\documentclass[12pt,a4paper]{article}
\usepackage{amssymb}
\usepackage{amsfonts}
\usepackage{amsmath}
\usepackage{latexsym}
\usepackage{epsfig}
\usepackage{epsf}
\usepackage{rotating}
\usepackage{url}
\usepackage{multicol}
\usepackage{cite}
\usepackage{physics}
\usepackage{xcolor}
\begin{document}

\title{\begin{flushright}
{\small INR-TH-2018-023}
\end{flushright} {\bf Classical behaviour of Q-balls\\ in the Wick--Cutkosky model}}

\author{
A.G.~Panin$^{a,b}$, M.N.~Smolyakov$^{c,a}$
\\
$^a${\small{\em
Institute for Nuclear Research of the Russian Academy
of Sciences,}}\\
{\small{\em 60th October Anniversary prospect 7a, Moscow 117312,
Russia
}}\\
$^b${\small{\em Moscow Institute of Physics and Technology,}}\\
{\small{\em Institutsky per. 9, Dolgoprudny 141700, Russia}}\\
$^c${\small{\em Skobeltsyn Institute of Nuclear Physics, Lomonosov Moscow
State University,
}}\\
{\small{\em Moscow 119991, Russia}}}

\date{}
\maketitle

\begin{abstract}
In this paper, we continue discussing Q-balls in the Wick--Cutkosky model. Despite Q-balls in this model are composed of two scalar fields, they turn out to be very useful and illustrative for examining various important properties of Q-balls. In particular, in the present paper we study in detail (analytically and numerically) the problem of classical stability of Q-balls, including the nonlinear evolution of classically unstable Q-balls, as well as the behaviour of Q-balls in external fields in the non-relativistic limit.
\end{abstract}

\section{Introduction}
Among various models admitting solutions of the Q-ball \cite{Rosen0,Coleman:1985ki} type,\footnote{It is necessary to mention that solutions of the Q-ball type were discussed in the literature even earlier, see, for example, \cite{Finkelstein:1951zz,RosenRosenstock,Glasko}.} there exists a class of models with two different scalar fields (the complex one and the real one), which also admit the existence of the Q-ball--like solutions. The most known example of such a Q-ball is presented in the well-known paper \cite{Friedberg:1976me}. One can also recall a simplified version of the model of \cite{Friedberg:1976me}, in which the potential of the real scalar field is neglected \cite{Levin:2010gp} (see also \cite{Loiko:2018mhb}). In paper \cite{Nugaev:2016uqd} the simplest example of the two-field Q-ball was presented: instead of the quartic interaction of the scalar fields of \cite{Levin:2010gp}, the triple Yukawa interaction between the scalar fields, resulting in the well-known Wick–-Cutkosky model \cite{Wick,Cutkosky}, was considered. A remarkable feature of the model is that the main characteristics of Q-balls can be obtained analytically, including the form of the energy--charge dependence. The latter simplifies the subsequent analysis of the model.

Despite Q-balls in this model are composed of two scalar fields, the Wick--Cutkosky model turns out to be a simple and useful toy model for examining various properties of Q-balls and for testing the methods, which can be applied to Q-balls in other theories (including theories with the standard one-field Q-balls). A part of the analysis was performed in paper \cite{Nugaev:2016uqd}, in the present paper we focus on the problem of classical stability of Q-balls and on the behaviour of Q-balls in external fields.

It should be noted that the Schr\"{o}dinger--Poisson system, which appears, for example, when one considers the Newtonian limit for the Bose stars made of scalar fields \cite{Ruffini:1969qy,Seidel:1990jh,Marsh:2015wka}, is very similar to the Wick--Cutkosky model. Thus, we think that the results obtained for the case of the Wick--Cutkosky model can be useful for examining the Bose stars too.

First, we will examine the classical stability of Q-balls in the Wick–-Cutkosky model. The criterion of classical stability of two-field Q-balls was established in paper \cite{Friedberg:1976me}. The proof presented in that paper is based on examining the properties of the energy functional of the system, while keeping the charge fixed. An alternative proof along the lines of the Vakhitov-Kolokolov method \cite{VK,Kolokolov}, which is based on the use of only the linearized equations of motion, can be found in \cite{Panin:2016ooo}. The classical stability criterion states that a Q-ball is stable with respect to small perturbations (classically stable) if $\frac{dQ}{d\omega}<0$ for this Q-ball and if a certain operator (we will denote this operator as $L_{+}$), which arises in the linearized equations of motion for perturbations, has only one negative eigenvalue. Usually, the operator $L_{+}$ indeed has only one negative eigenvalue. However, it is not a rule in the general case (except for the standard one-field Q-balls in (1+1)-dimensional space-time), so in any particular case one should check that the number of negative eigenvalues of the operator $L_{+}$ does not exceed one in order to be sure that Q-balls with $\frac{dQ}{d\omega}<0$ are classically stable. Below we will show explicitly that the latter condition holds for Q-balls in the Wick--Cutkosky model.

Second, for the case of classically unstable Q-balls, we will find explicitly the exponentially growing (instability) mode which is a solution to the corresponding linearized equations of motion. Next, we will examine the nonlinear evolution of classically unstable Q-balls by simulating numerically the evolution of the perturbed Q-ball. We will show that, depending on the characteristics of the initial unstable Q-ball and on the contribution of the instability mode into the initial perturbation, unstable Q-balls in the Wick–-Cutkosky model can evolve in three completely different ways (analogous evolution of classically unstable Q-balls may occur in other models admitting both classically stable and classically unstable Q-balls): the first way ends up at a classically stable Q-ball with the charge different from the one of the initial unstable Q-ball, the second way results in spreading of the Q-ball into spherical waves, the third way corresponds to the spherically symmetric collapse of the Q-ball. The latter process is analogous to the nonlinear self-similar evolution observed in \cite{Levkov}, that forces the particles to fall into the Bose star center.

And third, we will briefly examine the behaviour of classically stable Q-balls in external fields. From the classical point of view, the most interesting cases are those of the long-range interactions. Thus, the most suitable example seems to be the one that is provided by the electromagnetic interaction. In particular, classical motion of the Q-ball with logarithmic scalar field potential \cite{Rosen1} in external electromagnetic fields was thoroughly examined in \cite{BF,BF2}. Meanwhile, it is clear that if a Q-ball interacts with some sort of the field, it turns out to be the source of this field. For the case of electromagnetic fields, we get U(1) gauged Q-balls \cite{Rosengauged,Lee:1988ag} instead of ordinary Q-balls (in particular, for the logarithmic scalar field potential we get the model of \cite{Dzhunushaliev:2012zb}). Although in some cases the backreaction of the U(1) gauge field on the scalar field can be neglected for U(1) gauged Q-balls \cite{Gulamov:2013cra}, the existence of the gauge field can considerably modify the Q-ball properties. In particular, the question about the classical stability of U(1) gauged Q-balls has not been solved yet. Moreover, it is possible (although it is not proved) that all U(1) gauged Q-balls are unstable with respect to non-spherically symmetric perturbations \cite{Panin:2016ooo}. In such a case, the interaction of a U(1) gauged Q-ball with the external electromagnetic field may eventually destroy the Q-ball.

From this point of view, the Wick--Cutkosky model is very attractive. On the one hand, one may expect that there exist classically stable Q-balls. On the other hand, the real scalar field in this model is massless, which ensures the existence of long-range interactions. Moreover, the back-reaction of the real scalar field can not be neglected --- it is the real scalar field that provides the attraction for the complex scalar field and ensures the Q-ball existence. The latter makes the analysis even more interesting. We will show that, as expected, the Q-ball as a whole indeed obeys the Newton law, at least in the case of small speeds and accelerations. Meanwhile, the accelerated motion of the Q-ball modifies its form in accordance with its speed and acceleration.

Now let us turn to the description of the Wick--Cutkosky model.

\section{The model}
\begin{figure}[!ht]
\centering
\includegraphics[width=0.95\linewidth]{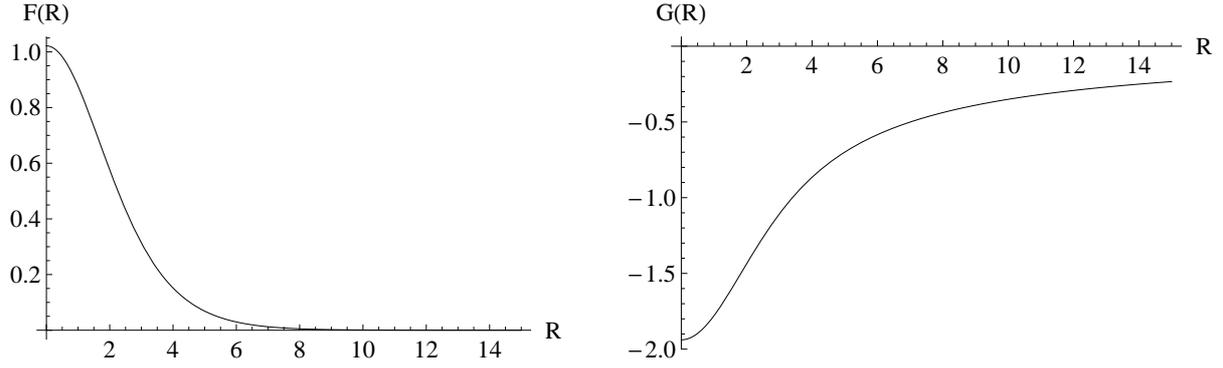}
\caption{Numerical solution for the fields $F(R)$ and $G(R)$ \cite{Nugaev:2016uqd}.}\label{solfig}
\end{figure}

The action of the Wick--Cutkosky model \cite{Wick,Cutkosky} has the form
\begin{equation}\label{action}
S=\int\left(\partial_\mu\chi^*\partial^\mu\chi+\frac{1}{2}\partial_\mu\tilde\phi\partial^\mu\tilde\phi-m^{2}\chi^*\chi-h\tilde\phi\chi^*\chi\right)d^4x,
\end{equation}
where $\tilde\phi\to 0$ for $|\vec x|\to\infty$, $m^{2}>0$, $h>0$. For the Q-ball, we take the standard ansatz
\begin{eqnarray}\label{chiequiv}
\chi(t,\vec x)=e^{i\omega t}f(r),\qquad \tilde\phi(t,\vec x)=g(r).
\end{eqnarray}
Here $r=\sqrt{{\vec x}^{2}}$ and $f(r)$ is a real function without nodes (without loss of generality, we can set $f(r)>0$ for any $r$). The functions $f(r)$ and $g(r)$ satisfy the boundary conditions
\begin{align}\label{boundary1}
\partial_{r}f|_{r=0}=0,\qquad \lim\limits_{r\to\infty}f(r)=0,\qquad
\partial_{r}g|_{r=0}=0, \qquad \lim\limits_{r\to\infty}g(r)=0.
\end{align}
From the very beginning, it is convenient to pass to the new variables
\begin{equation}
R=r\sqrt{m^{2}-\omega^{2}},\qquad F(R)=\frac{h}{m^2-\omega^2}f(r),\qquad G(R)=\frac{h}{m^2-\omega^2}g(r),
\label{transformation}
\end{equation}
for which the equations of motion, following from \eqref{action} with \eqref{chiequiv}, take the simple form
\begin{align}\label{eq1a}
&-\frac{1}{R}\partial_{R}^{2}(RF)+F+FG=0,\\ \label{eq2a}
&-\frac{1}{R}\partial_{R}^{2}(RG)+F^2=0
\end{align}
with the boundary conditions
\begin{equation}\label{boundary1a}
\partial_{R}F|_{R=0}=0,\qquad \lim\limits_{R\to\infty}F=0,\qquad
\partial_{R}G|_{R=0}=0, \qquad \lim\limits_{R\to\infty}G=0.
\end{equation}
One can see that there is no dependence on the frequency $\omega$ in these equations. The numerical solution to equations \eqref{eq1a}, \eqref{eq2a}, which was found in \cite{Nugaev:2016uqd}, is presented in Fig.~\ref{solfig}.

\begin{figure}[!h]
\centering
\includegraphics[width=0.5\linewidth]{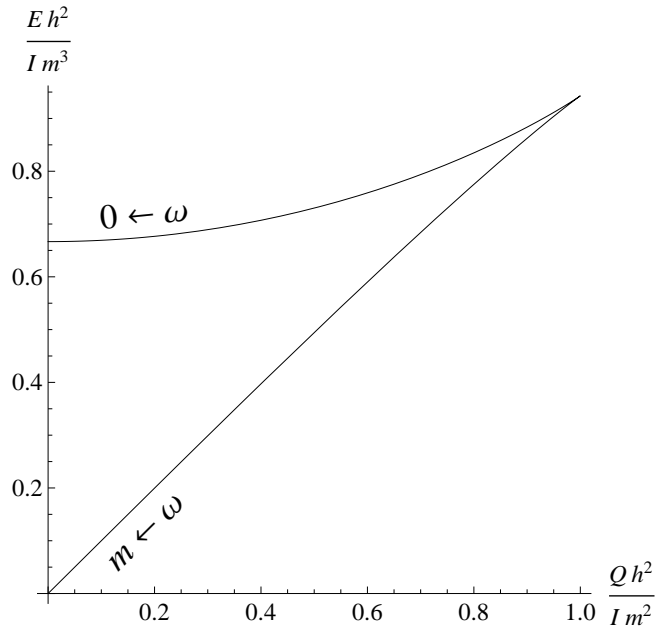}
\caption{$E(Q)$ diagram for $Q>0$ \cite{Nugaev:2016uqd}. The cusp corresponds to $\omega=\frac{m}{\sqrt{2}}$.}\label{eqfig}
\end{figure}

The charge and the energy of the Q-ball take the form \cite{Nugaev:2016uqd}
\begin{equation}\label{Q}
Q=\frac{2I}{h^2}\,\omega\sqrt{m^2-\omega^{2}},\qquad
E=\frac{I}{h^2}\sqrt{m^2-\omega^{2}}\left(\frac{4}{3}\omega^{2}+\frac{2}{3}m^{2}\right),
\end{equation}
where
\begin{equation}
I=4\pi\int\limits_{0}^{\infty}F^{2}R^{2}dR\approx 44.05.
\end{equation}
Due to the symmetry $\omega\to-\omega$ $\Longrightarrow$ $Q\to -Q$, $E\to E$, without loss of generality from here on we will consider the case $Q\ge 0$, which corresponds to $0\le\omega<m$, see Fig.~\ref{eqfig}.

\begin{figure}[!ht]
\centering
\includegraphics[width=0.95\linewidth]{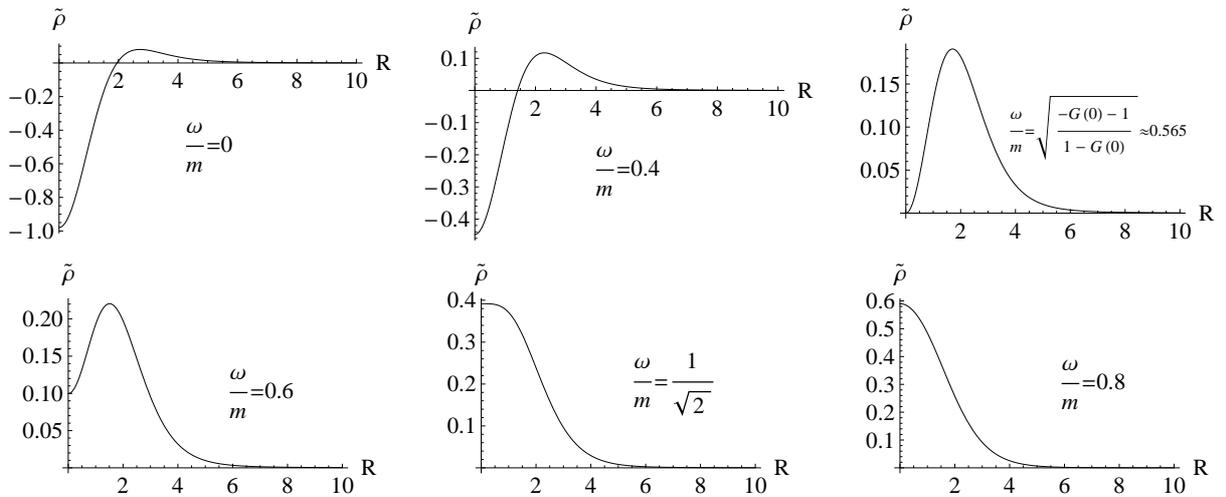}
\caption{Dimensionless energy density of the Q-ball for different values of $\omega$.}\label{figureED}
\end{figure}

It is illustrative to look at the dimensionless energy density $\tilde\rho(R)$ of the Q-ball, which is defined as
\begin{eqnarray}\nonumber
E=\frac{m^{3}}{h^{2}}4\pi\int\limits_{0}^{\infty}\tilde\rho(R)R^{2}dR\\=\frac{m^{3}}{h^{2}}4\pi\left(1-\frac{\omega^{2}}{m^{2}}\right)^{\frac{3}{2}}\int
\limits_{0}^{\infty}\left(\left(\partial_{R}F\right)^{2}+\frac{1}{2}\left(\partial_{R}G\right)^{2}+G F^{2}+\frac{m^{2}+\omega^{2}}{m^{2}-\omega^{2}}F^{2}\right)R^{2}dR.
\end{eqnarray}
The plots of the Q-ball dimensionless energy density for different values of $\omega$ are presented in Fig.~\ref{figureED}.
We see that there exist areas with negative energy density in the Q-ball center for $\left|\frac{\omega}{m}\right|<\sqrt{\frac{-G(0)-1}{1-G(0)}}\approx 0.565$. Analogous areas with negative energy density were observed in \cite{Anderson:1970et} for ordinary one-field Q-balls in a theory with the scalar field potential $V(\chi^{*}\chi)=m^{2}\chi^{*}\chi-\kappa\left(\chi^{*}\chi\right)^{2}$, $\kappa>0$.\footnote{It is easy to prove that at least in the case of the standard one-field Q-balls, the energy density at $r=0$ is such that $\rho(0)<0$, $\frac{d^{2}\rho(r)}{dr^{2}}\bigl|_{r=0}>0$ for $\omega=0$ and, consequently, for the values of $\omega$ in some vicinity of the point $\omega=0$; see Appendix for details.}

\section{Classical (in)stability}
Now we turn to examining the classical stability of Q-balls in the Wick--Cutkosky model. We will discuss two aspects of this problem: stability with respect to small perturbations and nonlinear evolution of the classically unstable Q-balls. Let us start with the first aspect.

\subsection{Stability with respect to small perturbations}\label{stab-analyt}
As was mentioned in Introduction, in order to be sure that Q-balls with $\frac{dQ}{d\omega}<0$ are classically stable, it is necessary to check that the operator $L_{+}$ has only one negative eigenvalue \cite{Panin:2016ooo}. The form of the operator $L_{+}$ in the Wick--Cutkosky model can be easily found using the results of paper \cite{Panin:2016ooo}:
\begin{equation}\label{Lplusedef}
L_{+}=
\begin{pmatrix}
-\Delta+m^{2}-\omega^{2}+hg(r) & hf(r) \\
hf(r) & -\frac{1}{2}\Delta
\end{pmatrix},
\end{equation}
where $\Delta=\sum\limits_{i=1}^{3}\partial_{i}^{2}$. For examining this operator, it is convenient to introduce the variables
\begin{equation}
\vec X=\vec x \sqrt{m^{2}-\omega^{2}}.
\end{equation}
In the notations $\vec X$, $F(R)$ and $G(R)$, where $R=\sqrt{\vec X^{2}}$ (see \eqref{transformation}), the corresponding eigenvalue problem for the operator $L_{+}$ takes the form
\begin{equation}\label{Leq}
\begin{pmatrix}
-\Delta_{X}+1+G(R) & F(R) \\
F(R) & -\frac{1}{2}\Delta_{X}
\end{pmatrix}\Psi_{N}(\vec X)=\Lambda_{N}\Psi_{N}(\vec X).
\end{equation}
where $\Delta_{X}=\sum\limits_{i=1}^{3}\frac{\partial^{2}}{{\partial X^{i}}^{2}}$. There is no dependence on $\omega$ in equation \eqref{Leq}, which simplifies the subsequent analysis considerably. The eigenfunction $\Psi_{N}(\vec X)$ can be represented in spherical coordinates as
\begin{equation}\label{nonsps}
\Psi_{N}(\vec X)=\Psi_{n}^{l}(R)Y_{lm}(\theta,\varphi)=\begin{pmatrix}
\xi_{n}^{l}(R)\\
\eta_{n}^{l}(R)
\end{pmatrix}Y_{lm}(\theta,\varphi),
\end{equation}
which results in
\begin{eqnarray}\label{lineq1}
&&-\frac{1}{R}\partial_{R}^{2}(R\xi_{n}^{l})+\frac{l(l+1)}{R^{2}}\xi_{n}^{l}+(1+G)\xi_{n}^{l}+F\eta_{n}^{l}=\Lambda_{n}^{l}\xi_{n}^{l},\\\label{lineq2}
&&-\frac{1}{2R}\partial_{R}^{2}(R\eta_{n}^{l})+\frac{l(l+1)}{2R^{2}}\eta_{n}^{l}+F\xi_{n}^{l}=\Lambda_{n}^{l}\eta_{n}^{l}
\end{eqnarray}
with the asymptotic $\xi_{n}^{l}(R)\sim R^{l},\, \eta_{n}^{l}(R)\sim R^{l}$ for $R\to 0$.

Now let us define the operator $L_{X}^{l}$ as
\begin{equation}
L_{X}^{l}=\begin{pmatrix}
-\frac{1}{R}\partial_{R}^{2}(R\,\cdot\,)+\frac{l(l+1)}{R^{2}}+(1+G(R)) & F(R) \\
F(R) & -\frac{1}{2R}\partial_{R}^{2}(R\,\cdot\,)+\frac{l(l+1)}{2R^{2}}
\end{pmatrix},
\end{equation}
leading to
\begin{equation}
L_{X}^{l+1}=L_{X}^{l}+\begin{pmatrix}
\frac{2(l+1)}{R^{2}} & 0 \\
0 & \frac{(l+1)}{R^{2}}
\end{pmatrix}.
\end{equation}
Let $\Psi_{0}^{l+1}(R)$ be the eigenfunction of the operator $L_{X}^{l+1}$, corresponding to the minimal eigenvalue $\Lambda_{0}^{l+1}$ such that $\bra{\Psi_{0}^{l+1}}\ket{\Psi_{0}^{l+1}}=1$. Since the operator $L_{X}^{l}$ is Hermitian, the function $\Psi_{0}^{l+1}(R)$ can be decomposed into eigenfunctions of the operator $L_{X}^{l}$ forming a complete orthonormal set:\footnote{Of course, there exists a continuous spectrum for $\Lambda_{N}>0$, so it is more appropriate to use $\sum\limits_{n=0}^{n_{d}}+\int\limits_{\Lambda_{c}}^{\infty}d\Lambda$ instead of $\sum\limits_{n=0}^{\infty}$. Moreover, the function $\Psi_{0}^{l+1}(R)$ can belong to the continuous spectrum, in such a case it should be normalized to delta-function, not to unity. However, the continuous spectrum can be easily transformed into a discrete spectrum by putting the system into a ``box'' of a finite size, whereas technically it is simpler to work with the discrete spectrum.}
\begin{equation}
\Psi_{0}^{l+1}(R)=\sum\limits_{n=0}^{\infty}c_{n}\Psi_{n}^{l}(R),\qquad \bra{\Psi_{n}^{l}}\ket{\Psi_{n}^{l}}=1,\qquad \sum\limits_{n=0}^{\infty}c_{n}^{2}=1.
\end{equation}

\begin{figure}[!ht]
\begin{tabular}{cc}
\includegraphics[width=0.47\linewidth]{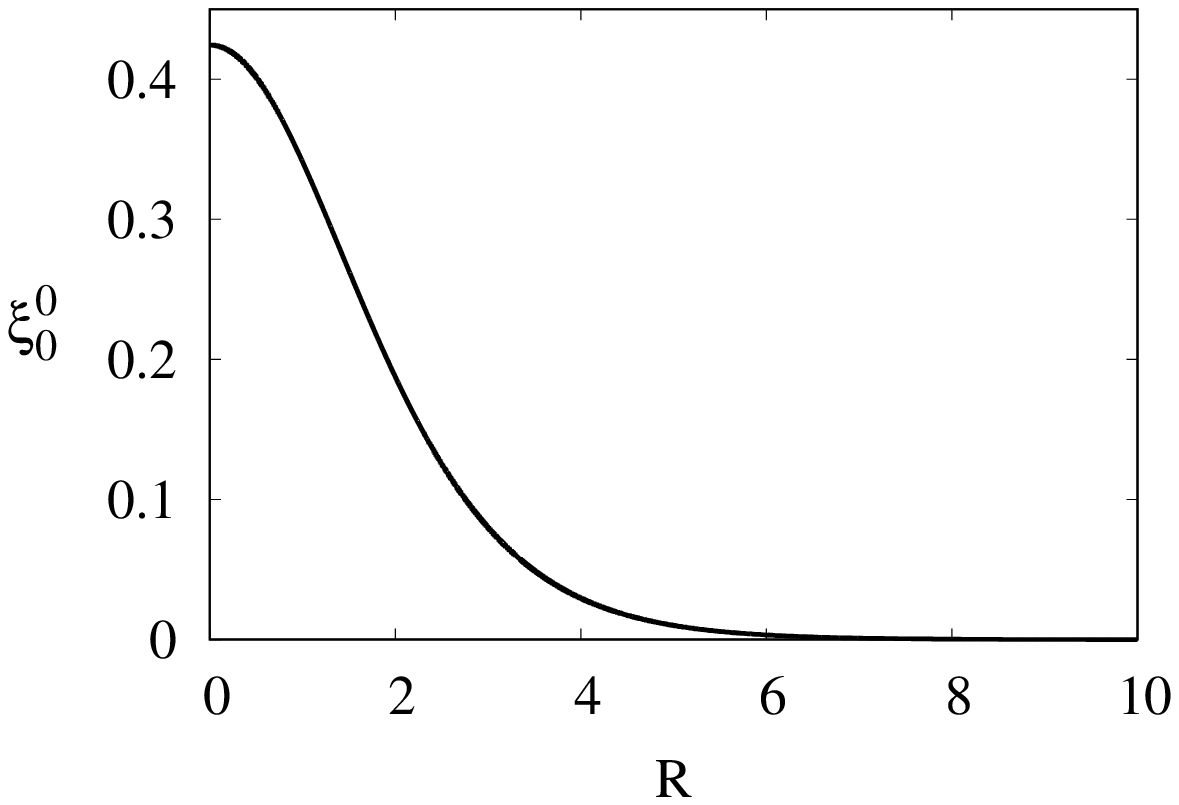}&
\includegraphics[width=0.47\linewidth]{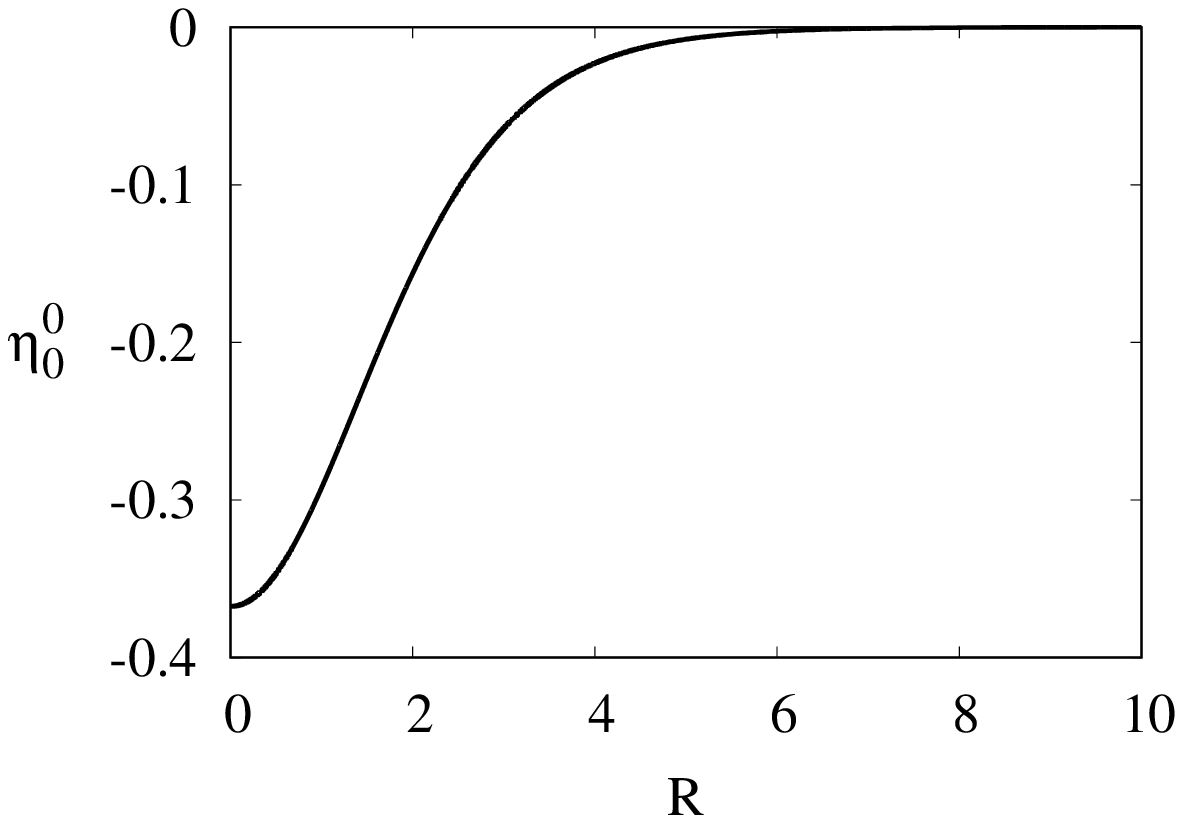}\\
\end{tabular}
\caption{The normalized eigenfunction corresponding to $\Lambda_{0}^{0}\approx -0.474$.}
\label{figure4}
\end{figure}

Now we can write for $\expval{L_{X}^{l+1}}{\Psi_{0}^{l+1}}=\Lambda_{0}^{l+1}$
\begin{eqnarray}\nonumber
\Lambda_{0}^{l+1}=\sum\limits_{n=0}^{\infty}c_{n}^{2}\Lambda_{n}^{l}+2(l+1)\int\limits_{0}^{\infty}\left(\xi_{0}^{l+1}(R)\right)^{2}dR+
(l+1)\int\limits_{0}^{\infty}\left(\eta_{0}^{l+1}(R)\right)^{2}dR>\sum\limits_{n=0}^{\infty}c_{n}^{2}\Lambda_{n}^{l}\\ \nonumber
=c_{0}^{2}\Lambda_{0}^{l}+\sum\limits_{n=1}^{\infty}c_{n}^{2}\Lambda_{n}^{l}=\left(1-\sum\limits_{n=1}^{\infty}c_{n}^{2}\right)\Lambda_{0}^{l}+\sum\limits_{n=1}^{\infty}c_{n}^{2}\Lambda_{n}^{l}
=\Lambda_{0}^{l}+\sum\limits_{n=1}^{\infty}c_{n}^{2}\left(\Lambda_{n}^{l}-\Lambda_{0}^{l}\right).
\end{eqnarray}
Since by the definition $\Lambda_{n}^{l}\ge\Lambda_{0}^{l}$ for $n\ge 1$, then
\begin{equation}
\Lambda_{0}^{l+1}>\Lambda_{0}^{l}.
\end{equation}
Now recall that there exists an exact solution to equations \eqref{lineq1}, \eqref{lineq2}. It has the form
\begin{equation}\label{transmode}
\xi_{0}^{1}=C\partial_{R}F,\qquad \eta_{0}^{1}=C\partial_{R}G,\qquad \Lambda_{0}^{1}=0,
\end{equation}
where $C$ is a constant. This solution corresponds to the three translational modes of the Q-ball. It is expected that this solution has the smallest eigenvalue for $l=1$. Thus, it is sufficient to check that eigenfunction \eqref{transmode} has the smallest eigenvalue among the eigenfunctions with $l=1$ and to look for the spherically symmetric ($l=0$) eigenfunctions with negative eigenvalues.\footnote{Analogous simplification of the analysis of the operator $L_{+}$ can be used in other models of Q-balls.} For problem \eqref{lineq1}, \eqref{lineq2}, this was done numerically, revealing that solution \eqref{transmode} is indeed the eigenfunction of the lowest eigenstate with $l=1$, whereas there exists only one spherically symmetric eigenfunction with negative eigenvalue $\Lambda_{0}^{0}\approx -0.474$, see Fig.~\ref{figure4}.

Thus, since the operator $L_{+}$ in the Wick--Cutkosky model has only one negative eigenvalue, we can be sure that Q-balls with $\frac{dQ}{d\omega}<0$, i.e., Q-balls with $|\omega|>\frac{m}{\sqrt{2}}$ (see \eqref{Q}), are classically stable.

\subsection{Exponentially growing mode}
In order to find explicitly the instability mode for Q-balls with $\frac{dQ}{d\omega}>0$, let us consider the linearized equations of motion for perturbations against the Q-ball. As was shown in \cite{Panin:2016ooo}, the instability mode can be only of the form
\begin{align}\label{pertdef2fields}
&\chi(t,\vec x)=e^{i\omega t}f(r)+e^{i\omega t}e^{\gamma t}\left(u(\vec x)+iv(\vec x)\right),\\ \label{pertchidef2fields}
&\tilde\phi(t,\vec x)=g(r)+e^{\gamma t}\varphi(\vec x),
\end{align}
where $u(\vec x)$, $v(\vec x)$ and $\varphi(\vec x)$ are real functions. The corresponding linearized equations of motion take the form \cite{Panin:2016ooo}
\begin{align}
&-\Delta u+(m^{2}-\omega^{2}+hg)u+hf\varphi=-\gamma^{2}u+2\omega\gamma v,\\
&-\frac{1}{2}\Delta\varphi+hfu=-\frac{1}{2}\gamma^{2}\varphi,\\
&-\Delta v+(m^{2}-\omega^{2}+hg)v=-\gamma^{2}v-2\omega\gamma u.
\end{align}
It is clear that the fields $u(\vec x)$, $v(\vec x)$ and $\varphi(\vec x)$ can be represented in spherical coordinates and decomposed as
\begin{equation}
u(\vec x)=u_{sp}(r)+u_{nsp}(\vec x),\qquad v(\vec x)=v_{sp}(r)+v_{nsp}(\vec x),\qquad \varphi(\vec x)=\varphi_{sp}(r)+\varphi_{nsp}(\vec x),
\end{equation}
where $u_{sp}(r)$, $v_{sp}(r)$ and $\varphi_{sp}(r)$ are spherically symmetric, whereas $u_{nsp}(\vec x)$, $v_{nsp}(\vec x)$ and $\varphi_{nsp}(\vec x)$ are non-spherically symmetric (i.e., include all the modes with the nonzero momenta $l$ in spherical coordinates). The equations of motion for the spherically symmetric parts of the fields and for the non-spherically symmetric parts of the fields decouple. In particular, for $u_{nsp}(\vec x)$, $v_{nsp}(\vec x)$ and $\varphi_{nsp}(\vec x)$ we get
\begin{align}\label{nspeq1}
&L_{+}\begin{pmatrix}
u_{nsp} \\ \varphi_{nsp}
\end{pmatrix}=-\gamma^{2}\begin{pmatrix}
u_{nsp} \\ \frac{1}{2}\varphi_{nsp}
\end{pmatrix}+2\omega\gamma\begin{pmatrix}
v_{nsp} \\ 0 \end{pmatrix},\\\label{nspeq2}
&L_{v}v_{nsp}=-\gamma^{2}v_{nsp}-2\omega\gamma u_{nsp},
\end{align}
where $L_{v}=-\Delta+(m^{2}-\omega^{2}+hg)$. Now let us multiply equation \eqref{nspeq1} by $\begin{pmatrix}
u_{nsp} & \varphi_{nsp}\end{pmatrix}$ and integrate the result over the space. Then, let us multiply equation \eqref{nspeq2} by $v_{nsp}$ and integrate the result over the space. Combining the resulting expressions, we can obtain
\begin{equation}\label{nspeq3}
\int\begin{pmatrix}
u_{nsp} & \varphi_{nsp}\end{pmatrix}L_{+}\begin{pmatrix}
u_{nsp} \\ \varphi_{nsp}
\end{pmatrix}d^{3}x+\int v_{nsp}L_{v} v_{nsp} d^{3}x=-\gamma^{2}\int\left(
u_{nsp}^{2}+\frac{1}{2}\varphi_{nsp}^{2}+v_{nsp}^{2}\right)d^{3}x.
\end{equation}
Recall that there is only one negative eigenvalue of the operator $L_{+}$, the corresponding eigenfunction is spherically symmetric. Thus, the first integral in \eqref{nspeq3} is nonnegative. Next, recall that $L_{v}f=0$, which is just the equation for the Q-ball profile $f(r)$. Since the function $f(r)$ has no nodes, it corresponds to the lowest eigenvalue of the operator $L_{v}$, which is equal to zero. Thus, the second integral in \eqref{nspeq3} is also nonnegative (and it is equal to zero only for $v_{nsp}\equiv 0$). Consequently, relation \eqref{nspeq3} cannot be fulfilled for $\gamma^{2}>0$, which means that instability mode can be only spherically symmetric.

\begin{figure}[!ht]
\begin{tabular}{cc}
\includegraphics[width=0.47\linewidth]{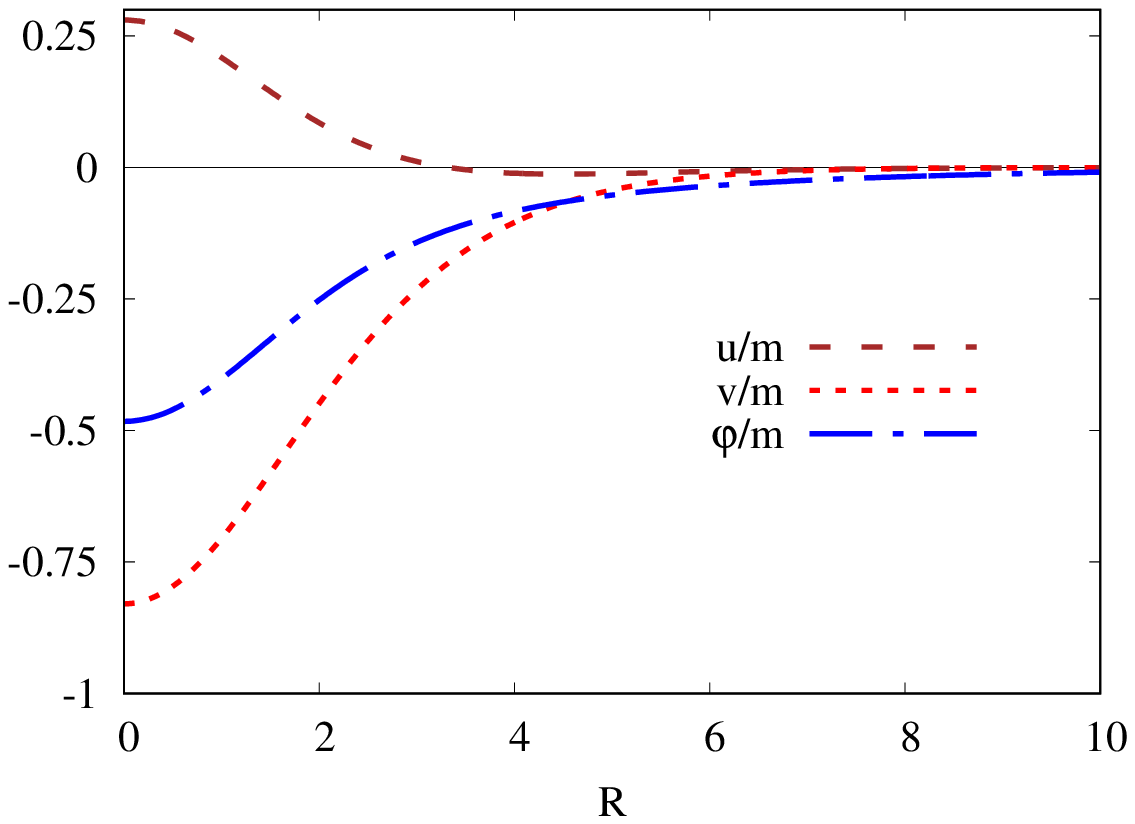}&
\includegraphics[width=0.47\linewidth]{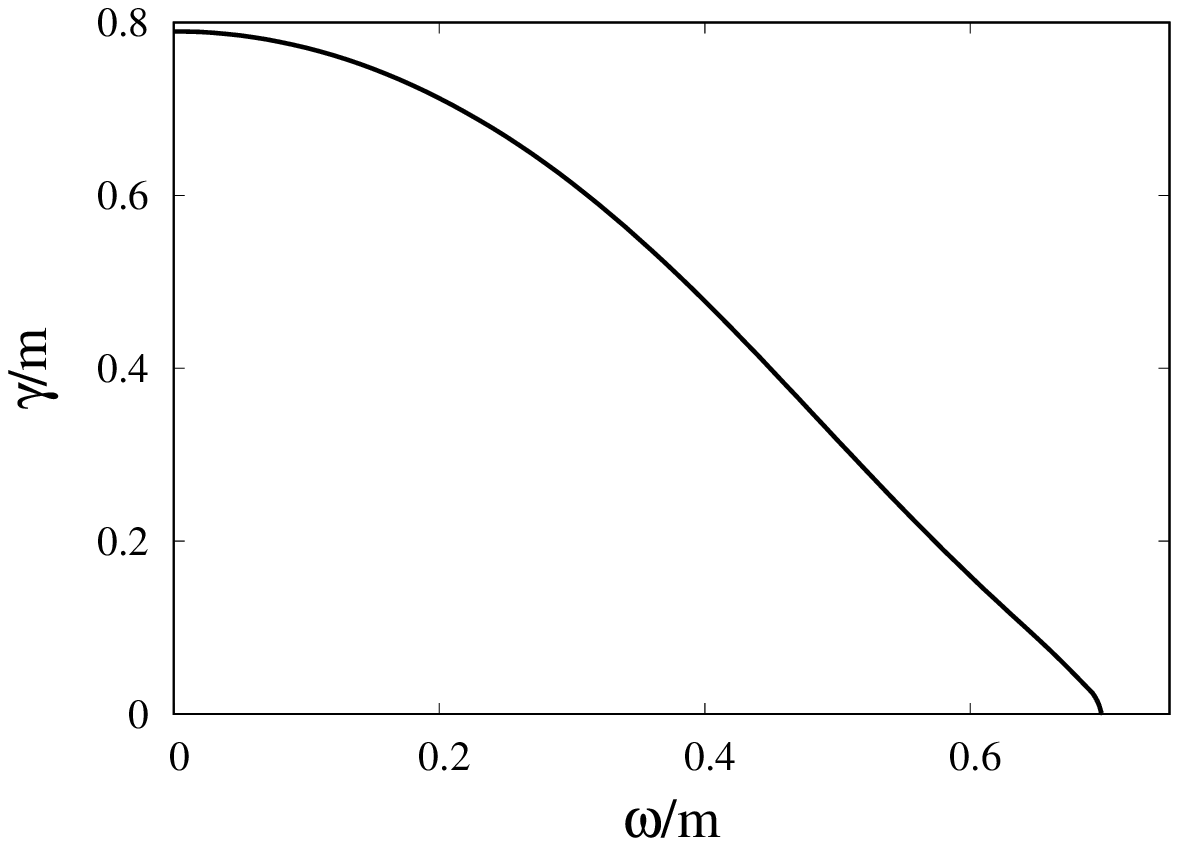}\\
\end{tabular}
\caption{The form of the instability mode for $\frac{\omega}{m}=0.6$ with the normalization $\frac{u^{2}(0)}{m^{2}}+\frac{v^{2}(0)}{m^{2}}+\frac{g^{2}(0)}{m^{2}}=1$ (left plot) and $\gamma(\omega)$ dependence (right plot).}
\label{figure5}
\end{figure}

Usually, there exists only one exponentially growing mode in the spectrum of excitations against a Q-ball.\footnote{Although we have not rigorous mathematical proof of this statement, we are not aware of any exceptions.} In order to find this instability mode explicitly, it is convenient to pass to variables \eqref{transformation}:
\begin{align}\label{instmodeeq1}
&-\frac{1}{R}\partial_{R}^{2}(Ru)+(1+G)u+F\varphi=-\frac{1}{1-\Omega^{2}}{\tilde\gamma}^{2}u+\frac{2\,\Omega}{1-\Omega^{2}}\tilde\gamma v,\\
&-\frac{1}{2R}\partial_{R}^{2}(R\varphi)+Fu=-\frac{1}{2(1-\Omega^{2})}{\tilde\gamma}^{2}\varphi,\\\label{instmodeeq3}
&-\frac{1}{R}\partial_{R}^{2}(Rv)+(1+G)v=-\frac{1}{1-\Omega^{2}}{\tilde\gamma}^{2}v- \frac{2\,\Omega}{1-\Omega^{2}}\tilde\gamma u,
\end{align}
where $\Omega=\frac{\omega}{m}$, $\tilde\gamma=\frac{\gamma}{m}$. Solutions to equations \eqref{instmodeeq1}--\eqref{instmodeeq3} were found numerically, some of the results are presented in Fig.~\ref{figure5}.

\subsection{Nonlinear evolution of the classically unstable Q-balls}
Now we turn to examining the nonlinear evolution of classically unstable Q-balls in the Wick--Cutkosky model, i.e., Q-balls with $|\omega|<\frac{m}{\sqrt{2}}$. We will consider only the spherically symmetric evolution. In this case, it is convenient to use the variables
\begin{equation}\label{defTROm}
R=r\sqrt{m^{2}-\omega^{2}},\qquad T=t\sqrt{m^{2}-\omega^{2}},\qquad \Omega=\frac{\omega}{m},
\end{equation}
\begin{equation}
Y(T,R)=\frac{h}{m^{2}-\omega^{2}}\chi(t,r),\qquad \Phi(T,R)=\frac{h}{m^{2}-\omega^{2}}\tilde\phi(t,r).
\end{equation}
In these variables, the nonlinear equations of motion take the form
\begin{eqnarray}\label{eqevol1}
&&\partial_{T}^{2}Y-\frac{1}{R}\partial_{R}^{2}(RY)+\frac{1}{1-\Omega^{2}}Y+Y\Phi=0,\\\label{eqevol2}
&&\partial_{T}^{2}\Phi-\frac{1}{R}\partial_{R}^{2}(R\Phi)+Y^{*}Y=0.
\end{eqnarray}
Here $-1<\Omega<1$. It is clear that it is sufficient to examine only the case $0\le\Omega<1$. For the stationary Q-ball solution, we have
\begin{equation}
Y(T,R)=e^{i\frac{\Omega}{\sqrt{1-\Omega^{2}}}T}F(R),\qquad \Phi(T,R)=G(R).
\end{equation}
Then we slightly perturb the stationary solution for the fields as
\begin{align}\label{pert1}
&Y(T,R)|_{T=0}=F(R)+\delta F(R),\qquad \partial_{T}Y(T,R)|_{T=0}=i\frac{\Omega}{\sqrt{1-\Omega^{2}}}F(R)+\delta{\dot F}(R),\\\label{pert2}
&\Phi(T,R)|_{T=0}=G(R)+\delta G(R),\qquad \partial_{T}\Phi(T,R)|_{T=0}=\delta{\dot G}(R).
\end{align}
$\delta F(R)$, $\delta{\dot F}(R)$, $\delta G(R)$, $\delta{\dot G}(R)$ are small perturbations generated randomly in the Fourier space with the white noise spectrum. It turns out that for the classically unstable Q-balls, the only relevant part of the perturbation is the exponentially growing instability mode, i.e., only the contribution of this mode defines the subsequent evolution of the dynamical system. Thus, in such a case we can take $\delta F(R)\sim u(R)+iv(R)$, $\delta G(R)\sim\varphi(R)$, $\delta{\dot F}(R)\sim \left(i\frac{\Omega}{\sqrt{1-\Omega^{2}}}+\frac{\tilde\gamma}{\sqrt{1-\Omega^{2}}}\right)\delta F(R)$, $\delta{\dot G}(R)\sim\frac{\tilde\gamma}{\sqrt{1-\Omega^{2}}}\delta G(R)$. Then we numerically evolve these perturbed configurations forward in time. In the simulations, we use the stable second-order iterated Crank-Nicolson scheme \cite{CrankNicolson} according to its realization presented in \cite{Teukolsky:1999rm}. Outgoing spherical waves that leave the interaction region are removed with the help of the Kreiss-Oliger filter \cite{KreissOliger}. Actually, here we use exactly the same numerical method as the one that was used in \cite{Panin:2016ooo} for examining the classical stability of U(1) gauged Q-balls; the detailed description of the method can be found in Appendix~C of \cite{Panin:2016ooo}.

The results of the numerical simulations show that, as expected, classically stable Q-balls (Q-balls with $\Omega>\frac{1}{\sqrt{2}}$) survive and drop some charge outside its core by means of spherical waves. The case of classically unstable Q-balls (Q-balls with $0\le\Omega<\frac{1}{\sqrt{2}}$) turns out to be more involved: the perturbations eventually destroy these Q-balls, but they can evolve in three completely different ways, depending on the form of the perturbation of the initial Q-ball and its parameters. If $\textrm{Re}\,\delta F(0)>0$, then all such Q-balls collapse. If $\textrm{Re}\,\delta F(0)<0$, then such Q-balls evolve in two different ways depending on the frequency of the initial Q-ball. Namely, if the frequency of the initial Q-ball is such that $0<\Omega_{in}\lesssim 0.38$ (here $\Omega_{in}$ is the frequency of the initial classically unstable Q-ball, see equations \eqref{pert1}--\eqref{pert2}), then such Q-balls simply spread into spherical waves;\footnote{It is necessary to mention that the Q-ball collapse (in finite time) and spreading into spherical waves were observed in \cite{Anderson:1970et}, where the nonlinear evolution of classically unstable Q-balls was examined numerically (note that all Q-balls in the model of \cite{Anderson:1970et} are classically unstable).} whereas initial Q-balls with $0.38\lesssim\Omega_{in}<\frac{1}{\sqrt{2}}$ drop some charge and turn into the classically stable Q-balls. Below we will discuss all these ways of the unstable Q-ball evolution in more detail.
\begin{figure}[ht]
\centering
\includegraphics[width=0.5\linewidth]{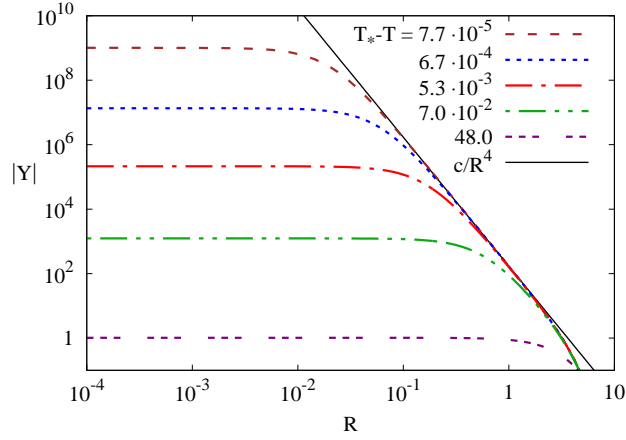}
\caption{Profile of the complex scalar field $Y(T,R)$ for $\Omega_{in}=0.6$ at different moments of time $T$. Here $c\approx 170$, $T_{*}\approx 48$.}
\label{figure6}
\end{figure}
\begin{figure}[!ht]
\begin{tabular}{cc}
\includegraphics[width=0.47\linewidth]{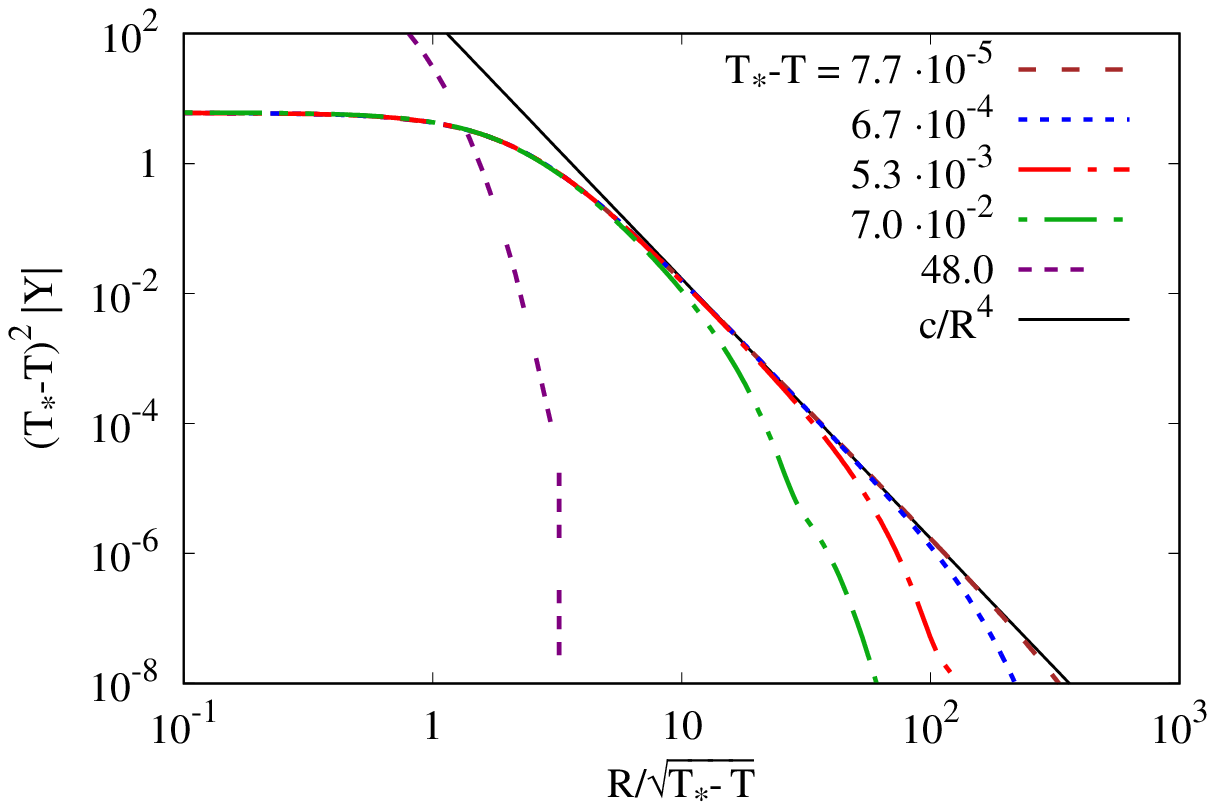}&
\includegraphics[width=0.47\linewidth]{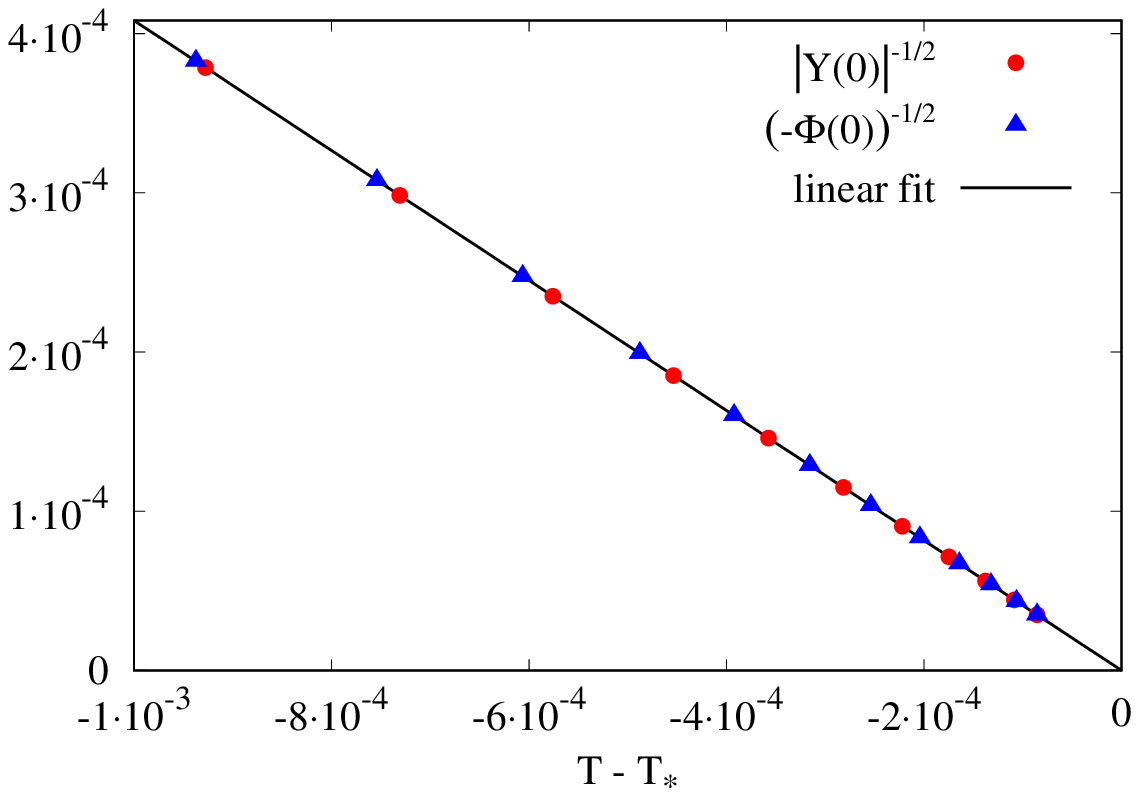}\\
\end{tabular}
\caption{Profile of the complex scalar field $Y(T,R)$ for $\Omega_{in}=0.6$ in the self-similar coordinates (left plot) and behaviour of the scalar fields $Y(T,R)$ and $\Phi(T,R)$ for $T\to T_{*}$ (right plot). Here $c\approx 170$, $T_{*}\approx 48$.}
\label{figure7}
\end{figure}

First, let us consider the case in which the contribution of the field $u$ of the instability mode into the perturbation is positive in the center of the initial Q-ball. In such a case, all classically unstable Q-balls collapse at finite time $T_{*}$.\footnote{The time $T_{*}$ depends not only on the characteristics of the initial classically unstable Q-balls, but also on the initial amplitude of the exponentially growing mode forming the perturbation. For example, for $\Omega_{in}=0.6$ and $\textrm{Re}\,\delta F(0)\approx 2.8\cdot 10^{-5}$, which corresponds to the mode of Fig.~\ref{figure5} multiplied by $10^{-4}$, the collapse time $T_{*}\approx 48$.} While approaching the collapse time $T_{*}$, the profile of the complex scalar field $Y(T,R)$ forms a central singularity $|Y|\sim\frac{1}{R^{4}}$, see Fig.~\ref{figure6}. It is interesting to note that the Q-ball collapse can be described by means of the self-similar coordinate $z=\frac{R}{\sqrt{T_{*}-T}}$. As can be seen from the left plot of Fig.~\ref{figure7}, the function $(T_{*}-T)^{2}|Y(T,R)|$ approaches some attractor function which depends on the self-similar coordinate $z$ only. An analogous self-similar evolution was described in \cite{Levkov} for the case of collapsing Bose stars, though with a different form of singularity (in the case of Bose stars, the form of singularity is $\sim\frac{1}{R}$ \cite{Levkov}). This function is not equal to zero at $z=0$, which means that $|Y(T,0)|\sim\frac{1}{(T_{*}-T)^{2}}$. More precisely, $|Y(T,0)|\approx\frac{6}{(T_{*}-T)^{2}}$, which corresponds to a solution of equations \eqref{eqevol1}, \eqref{eqevol2} with the terms $\frac{1}{R}\partial_{R}^{2}(RY)$, $\frac{1}{R}\partial_{R}^{2}(R\Phi)$ and $\frac{1}{1-\Omega^{2}}Y$ neglected. In this approximation, $\Phi\equiv -|Y|$. The latter is explicitly demonstrated on the right plot of Fig.~\ref{figure7}.

\begin{figure}[h]
\centering
\includegraphics[width=0.7\linewidth]{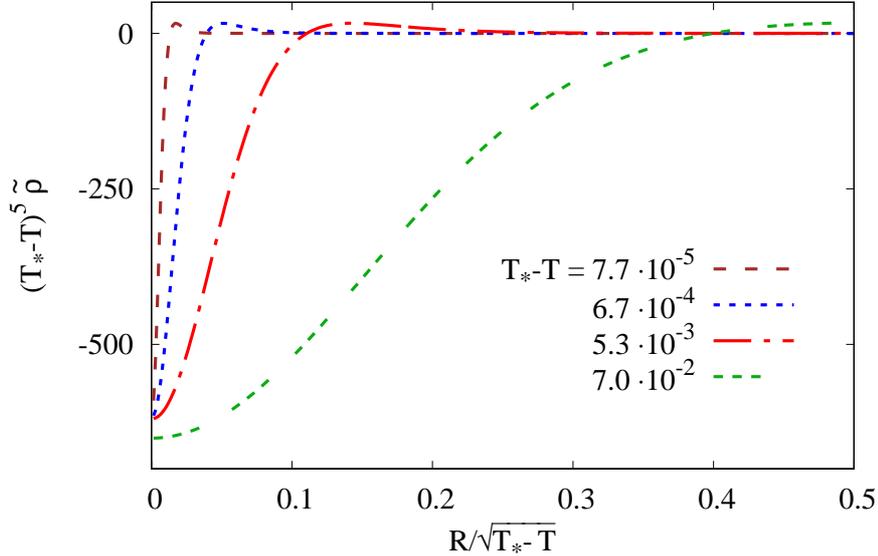}
\caption{Energy density of the collapsing Q-ball for $\Omega_{in}=0.6$ at different moments of time $T$. Here $T_{*}\approx 48$.}
\label{figure8}
\end{figure}

It is interesting to look at the energy density of the collapsing system. The dimensionless energy density $\tilde\rho(T,R)$ is defined as
\begin{eqnarray}\nonumber
E=\frac{m^{3}}{h^{2}}\left(1-\Omega^{2}\right)^{\frac{3}{2}}4\pi\int\limits_{0}^{\infty}\left(\partial_{T}Y^{*}\partial_{T}Y
+\partial_{R}Y^{*}\partial_{R}Y+\frac{1}{2}\left(\partial_{T}\Phi\right)^{2}+\frac{1}{2}\left(\partial_{R}\Phi\right)^{2}\right.\\\left.+\Phi Y^{*}Y+\frac{1}{1-\Omega^{2}}Y^{*}Y\right)R^{2}dR=\frac{m^{3}}{h^{2}}4\pi\int\limits_{0}^{\infty}\tilde\rho(T,R)R^{2}dR.\label{totEncoll}
\end{eqnarray}
The energy density also becomes singular as $T$ approaches $T_{*}$, the behaviour of the energy density of all collapsing Q-balls is similar regardless the value of the energy density at $R=0$, see the typical example in Fig.~\ref{figure8}.
One can see that $\lim\limits_{T\to T^{*}}\tilde\rho(T,0)=-\infty$, even though the initial energy density of the Q-ball with $\Omega_{in}=0.6$ at $R=0$ is positive (see Fig.~\ref{figureED}). This behaviour of the energy density is analogous to the one of the collapsing Q-ball in \cite{Anderson:1970et}. Meanwhile, since the total energy $E$ (see \eqref{totEncoll}) is conserved over time (it was checked numerically that it is conserved with a good accuracy during the collapse, which is an additional test of the correctness of the numerical simulation), it turns out that for any $R_{s}>0$ almost all energy of the initial Q-ball is concentrated within the sphere of radius $R_{s}$ at a fixed time $T_{s}<T_{*}$. One can choose this sphere to be of the Schwarzschild radius of the Q-ball or even smaller. In this connection, it is interesting to check whether or not the collapse of the classically unstable Q-balls leads to a production of non-rotating black holes along the lines of \cite{Choptuik:1992jv,Blinnikov:2016bxu,Helfer:2016ljl}. However, in order to check this possibility, it is necessary to perform numerical simulations taking into account the existence of the gravitational field. This problem calls for further detailed investigation.

\begin{figure}[!ht]
\centering
\includegraphics[width=0.7\linewidth]{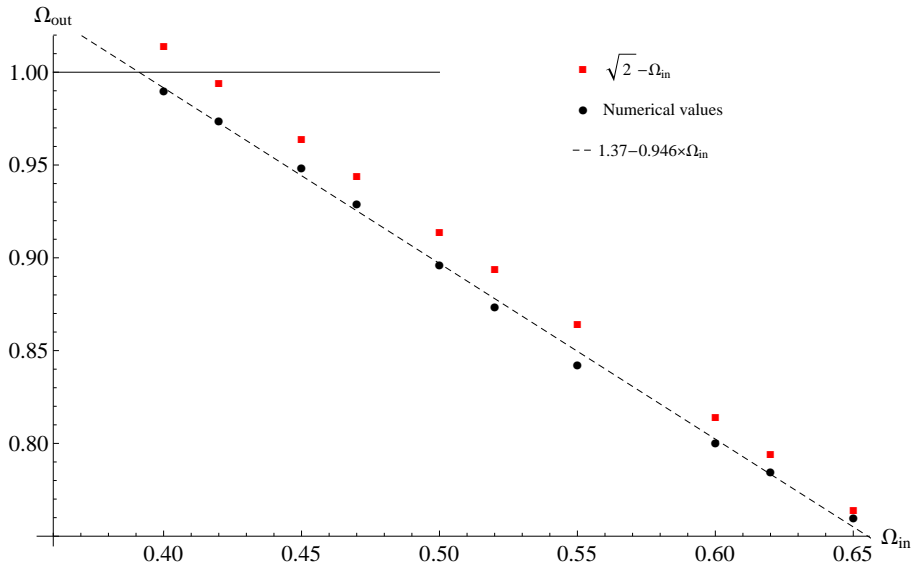}
\caption{The frequencies of the initial Q-balls $\Omega_{in}=\frac{\omega_{in}}{m}$ and the frequencies of the resulting Q-balls $\Omega_{out}=\frac{\omega_{out}}{m}$. The dashed line stands for the fit of the numerical data.}
\label{figure9}
\end{figure}

\begin{figure}[!!htb]
\begin{tabular}{cc}
\includegraphics[width=0.47\linewidth]{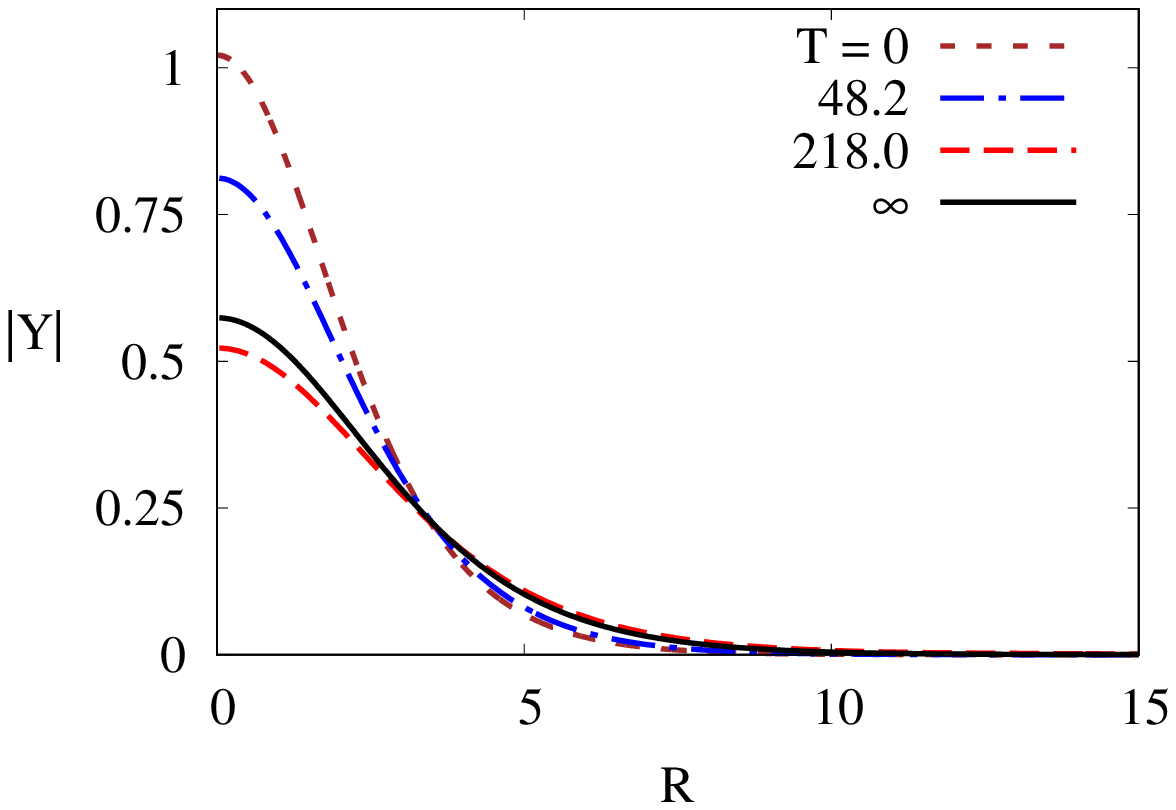}&
\includegraphics[width=0.47\linewidth]{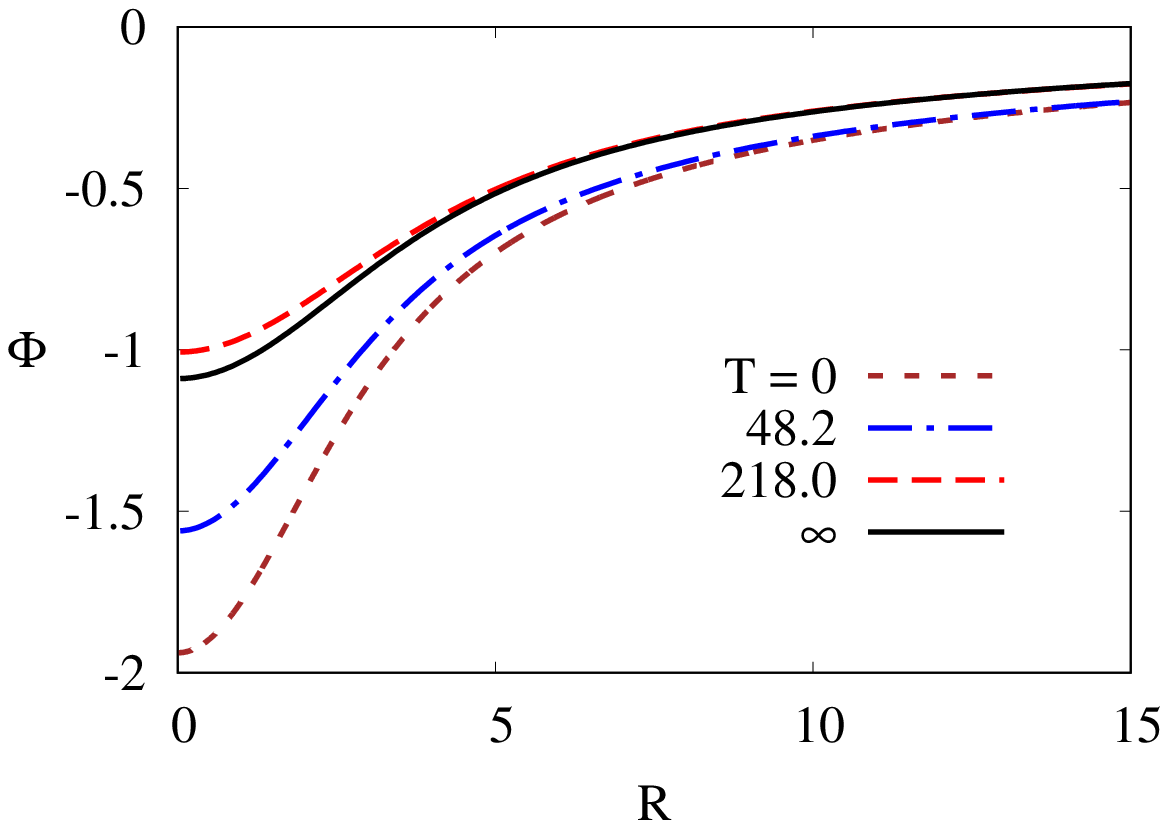}\\
\end{tabular}
\caption{Evolution of the Q-ball with $\Omega_{in}=0.6$.}
\label{figure10}
\end{figure}

\begin{figure}[!htb]
\centering
\includegraphics[width=0.6\linewidth]{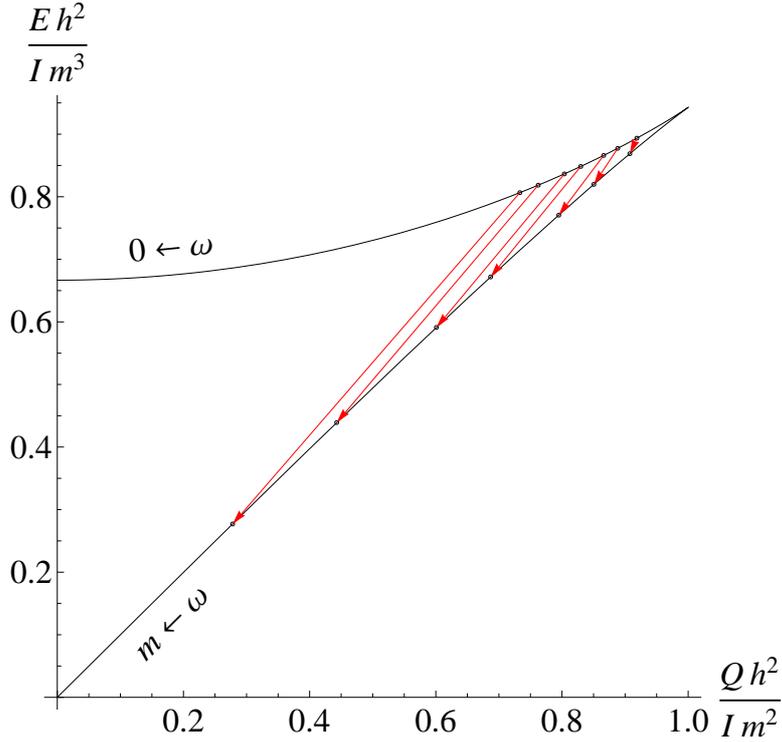}
\caption{Transitions from classically unstable Q-balls to classically stable Q-balls for $0.4\leq \Omega_{in}\leq 0.55$ (see Table~\ref{table}) on the $E(Q)$ diagram.}
\label{figure11}
\end{figure}

Now we turn to the case in which the contribution of the field $u$ of the instability mode into the perturbation is negative in the center of the initial Q-ball, i.e., $\textrm{Re}\,\delta F(0)<0$. As was noted above, in such a case there can be two different ways of the Q-ball evolution. The most interesting case is the one with $0.38\lesssim\Omega_{in}<\frac{1}{\sqrt{2}}$, in which classically unstable Q-balls transform into classically stable Q-balls. In Table~\ref{table}, the frequencies of the initial and resulting Q-balls are presented.
\begin{table}[ht]
    \begin{tabular}{| c | c | c | c | c | c | c | c | c | c | c |}
    \hline
    $\Omega_{in}$ & 0.4 & 0.42 & 0.45 & 0.47 & 0.5 & 0.52 & 0.55 & 0.6 & 0.62 & 0.65\\ \hline
    $\Omega_{out}$ & 0.9901 & 0.9738 & 0.9485 & 0.9292 & 0.8963 & 0.8735 & 0.8426 & 0.8004 & 0.7847 & 0.7600\\ \hline
    \end{tabular}
     \caption{The frequencies of the initial Q-balls $\Omega_{in}=\frac{\omega_{in}}{m}$ and the frequencies of the resulting Q-balls $\Omega_{out}=\frac{\omega_{out}}{m}$.}
     \label{table}
\end{table}
It turns out that the frequency of the resulting Q-ball can be roughly approximated by $\Omega_{out}\approx\sqrt{2}-\Omega_{in}$ (the fit of the numerical data results in $\Omega_{out}\approx 1.37-0.946\,\Omega_{in}$), see Fig.~\ref{figure9}.
An example of such an evolution for the Q-ball with $\Omega_{in}=0.6$ is presented in Fig.~\ref{figure10}.
An important point is that the charge of the resulting Q-ball is smaller (for some Q-balls it is much smaller) than the charge of the corresponding initial Q-ball, which contradicts the naive expectations that the charge of the resulting Q-ball is approximately the same as the charge of the initial Q-ball. The extra charge is carried away by spherical waves during the evolution. The transitions for Q-balls with $0.4\leq \Omega_{in}\leq 0.55$ (see Table~\ref{table}) are presented in Fig.~\ref{figure11}.

\begin{figure}[!htb]
\begin{tabular}{cc}
\includegraphics[width=0.47\linewidth]{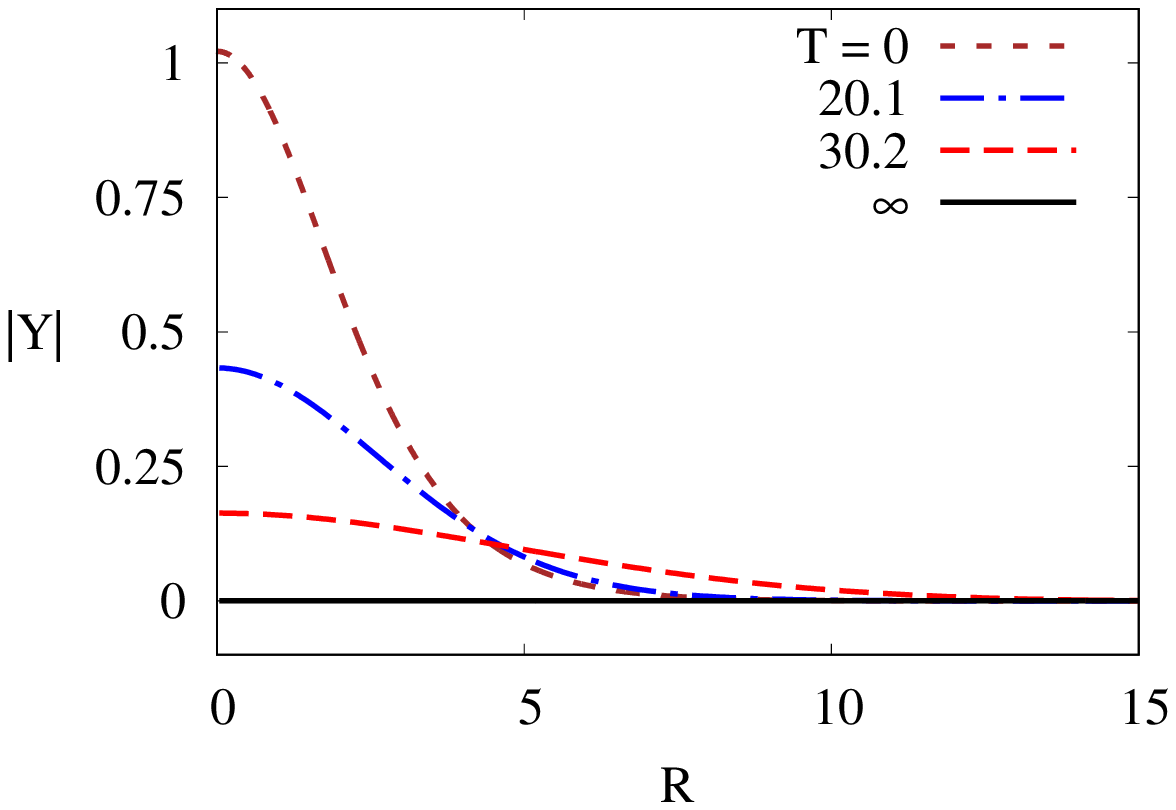}&
\includegraphics[width=0.47\linewidth]{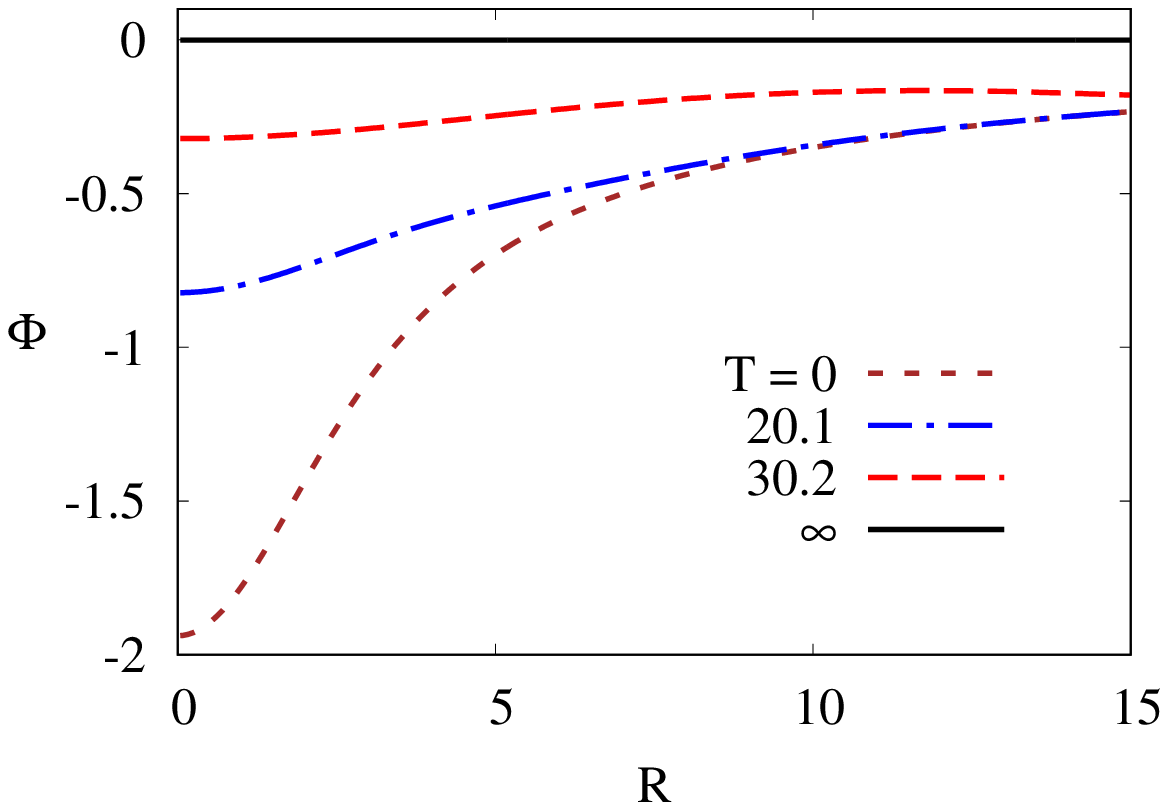}\\
\end{tabular}
\caption{Spreading of the Q-ball with $\Omega_{in}=0.37$.}
\label{figure12}
\end{figure}

As for the case $0<\Omega_{in}\lesssim 0.38$, all such Q-balls simply spread into spherical waves. An example of such a spreading is presented in Fig.~\ref{figure12}.

Now let us compare lifetimes of the Q-balls, which can be estimated with the help of the linearized theory discussed above, with the Q-ball decay times coming from the numerical simulations.\footnote{We are grateful to the anonymous referee of EPJC for suggesting us to perform this comparison.} In the linearized theory, the dimensionless lifetime can be estimated as $\tau_{lin}(\Omega)=\frac{1}{\tilde\gamma(\Omega)}$, where $\tilde\gamma(\Omega)=\frac{\gamma(\omega)}{m}$ is defined by the solutions to equations \eqref{instmodeeq1}--\eqref{instmodeeq3}. For example, for $\Omega_{in}=0.6$ we get $\tau_{lin}(0.6)\approx 6.3$. The collapse time $\tau_{coll}$ in the same units can be easily obtained from $T_{*}$ as $\tau_{coll}=\frac{T_{*}}{\sqrt{1-\Omega_{in}^{2}}}$ (see \eqref{defTROm}). For the case presented in Figs.~\ref{figure6}--\ref{figure8} with $T_{*}\approx 48$, we obtain $\tau_{coll}\approx 60$. Thus, we get
\begin{equation}\label{tau1rel}
\tau_{coll}\approx 9.5\,\tau_{lin}(0.6).
\end{equation}

For the spreading into spherical waves for $\Omega_{in}=0.37$ (Fig.~\ref{figure12}), we get $\tau_{lin}(0.37)\approx 1.9$, whereas the spreading time can be estimated as $\tau_{spr}=\frac{T_{spr}}{\sqrt{1-0.37^{2}}}\approx 41.4$ with $T_{spr}\approx 38.5$, leading to
\begin{equation}\label{tau2rel}
\tau_{spr}\approx 21.8\,\tau_{lin}(0.37).
\end{equation}
We define $T_{spr}$ such that $|Y(T_{spr},0)|=10^{-1}|Y(0,0)|$. In this case, $\Phi(T_{spr},0)\sim 10^{-1}\Phi(0,0)$, and for the energy density we get $\tilde\rho(T_{spr},0)\sim 10^{-2}\tilde\rho(0,0)$. In addition, according to the behaviour of $\frac{d|Y(T,0)|}{dT}$ obtained numerically, the spreading slows down considerably at times around $T_{spr}$. Thus, $T_{spr}$ seems to be a good choice for isolating the main stage of the decay, at least as an estimate.

For the transition from the classically unstable Q-ball to a classically stable Q-ball for $\Omega_{in}=0.6$ (Fig.~\ref{figure10}), we estimate the transition time as $\tau_{tr}=\frac{T_{tr}}{\sqrt{1-0.6^{2}}}\approx 100.6$ with $T_{tr}\approx 80.5$, leading to
\begin{equation}\label{tau3rel}
\tau_{tr}\approx 16\,\tau_{lin}(0.6).
\end{equation}
Here $T_{tr}$ corresponds to the smallest $T>0$ such that
\begin{equation}
\frac{d|Y(T,0)|}{dT}\biggl|_{T=T_{tr}}=0.
\end{equation}
For $T>T_{tr}$, the scalar fields oscillate (with damping) around the stable Q-ball configuration, and the system can be considered not as the decaying unstable Q-ball, but as the excited stable Q-ball.

Note that relations \eqref{tau1rel}--\eqref{tau3rel} were obtained for the particular choices of the initial perturbations. Different initial conditions may change the coefficients in \eqref{tau1rel}--\eqref{tau3rel}.

It is interesting to compare relations \eqref{tau1rel}--\eqref{tau3rel} with analogous relations for the Q-balls examined in \cite{Anderson:1970et}. Using the results presented in \cite{Anderson:1970et}, in the case $\omega'=0$ ($\omega'$ in \cite{Anderson:1970et} is analogous to $\Omega_{in}$) for the collapse (``singular decay'' in \cite{Anderson:1970et}) we get the estimate
\begin{equation}\label{tau4rel}
4\,\tau_{lin}(0)\lesssim\tau_{coll}\lesssim 10\,\tau_{lin}(0),
\end{equation}
and for the spreading into spherical waves (``dissipative decay'' in \cite{Anderson:1970et}) we get the estimate
\begin{equation}\label{tau5rel}
8\,\tau_{lin}(0)\lesssim\tau_{spr}\lesssim 16\,\tau_{lin}(0).
\end{equation}
The differences in the values of $\tau_{coll}$ and $\tau_{spr}$ in formulas \eqref{tau4rel} and \eqref{tau5rel} are caused by different forms of initial perturbations that were used in the numerical simulations. Analogously, in the case $\omega'=0.8$ we get from \cite{Anderson:1970et}
\begin{equation}\label{tau6rel}
\tau_{coll}\approx 5.3\,\tau_{lin}(0.8)
\end{equation}
and
\begin{equation}\label{tau7rel}
\tau_{spr}\approx 13.2\,\tau_{lin}(0.8).
\end{equation}
We see that although the model of \cite{Anderson:1970et} is different from the Wick--Cutkosky model, the coefficients in relations \eqref{tau4rel}--\eqref{tau7rel} are similar to those in \eqref{tau1rel}--\eqref{tau3rel}.

\section{Q-balls in external fields}
Finally, let us discuss how the classically stable Q-balls behave in the external long-range field which can be produced, say, by other Q-balls in the Wick--Cutkosky model. The equations of motion for the system in the presence of an external real scalar field, which is denoted by $U(\vec x)$, take the form
\begin{align}\label{eqtdep1}
&\ddot\chi-\Delta\chi+m^{2}\chi+h\chi\tilde\phi+h\chi U=0,\\ \label{eqtdep2}
&\ddot{\tilde\phi}-\Delta\tilde\phi+h\chi^{*}\chi=0,
\end{align}
where $\dot{}=\partial_{t}$. Of course, in the general case this system of equations can be solved only numerically. Thus, in order to get some analytical results, we suppose that $U(\vec x)\ll\textrm{max}|\tilde\phi(\vec x)|$, $hU(\vec x)\ll m^{2}$. Let us take the following ansatz for the scalar fields:
\begin{align}\label{subst1}
&\chi(t,\vec x)=e^{i\omega\left(t-(\dot{\vec R}\vec x)\right)}\left[f\left(\vec x-\vec R(t)\right)+u(t,\vec x)+iv(t,\vec x)\right],\\ \label{subst2}
&\tilde\phi(t,\vec x)=g\left(\vec x-\vec R(t)\right)+\rho(t,\vec x),
\end{align}
where $u(t,\vec x)$, $v(t,\vec x)$ and $\rho(t,\vec x)$ are real functions and
$$f\left(\vec x-\vec R(t)\right)=f\left(\sqrt{\left(\vec x-\vec R(t)\right)^{2}}\right),\qquad g\left(\vec x-\vec R(t)\right)=g\left(\sqrt{\left(\vec x-\vec R(t)\right)^{2}}\right)$$
is the Q-ball solution. Here $\vec R(t)$ denotes a position of the Q-ball center at the moment of time $t$. In what follows, we will consider the case $|\dot R_{i}|\ll 1$, $|\ddot R_{i}|/\omega\ll 1$. Then, the fields $u(t,\vec x)$, $v(t,\vec x)$ and $\rho(t,\vec x)$ can be assumed to be rather small to ensure the validity of the linear approximation for these fields. Now we substitute \eqref{subst1}, \eqref{subst2} into equations of motion \eqref{eqtdep1}, \eqref{eqtdep2} and retain the terms linear in $u$, $v$, $\rho$ (omitting the corrections containing $\dot R_{i}$, $\ddot R_{i}$ in the terms with $u$, $v$, $\rho$) and the terms $\sim\ddot R_{i}$, $\sim\dot R_{i}\dot R_{j}$. The terms containing more than two derivatives in time $t$ (like $\sim\ddot R_{i}\dot R_{j}$) are also omitted. We get the system of coupled equations
\begin{gather}\label{eq1motion}
\begin{pmatrix}
\ddot u-2\omega\dot v \\ \ddot\rho
\end{pmatrix}+L_{+}\begin{pmatrix}
u \\ \rho
\end{pmatrix}=\begin{pmatrix}
\sum\limits_{i=1}^{3}\left(\ddot{R_{i}}\partial_{i}f-2\omega^{2}f\ddot{R_{i}}x_{i}-\sum\limits_{j=1}^{3}
\dot{R_{i}}\dot{R_{j}}\partial_{i}\partial_{j}f-\omega^{2}f\dot{R_{i}}^{2}\right)-hf U
 \\ \frac{1}{2}\sum\limits_{i=1}^{3}\left(\ddot{R_{i}}\partial_{i}g-\sum\limits_{j=1}^{3}
\dot{R_{i}}\dot{R_{j}}\partial_{i}\partial_{j}g\right)
\end{pmatrix},
\\\label{eq2motion}
\ddot v+2\omega\dot u-\Delta v+(m^{2}-\omega^{2})v+hgv=0,
\end{gather}
where $\partial_{i}f=\frac{\partial f\left(\vec x-\vec R(t)\right)}{\partial x_{i}}$, $\partial_{i}g=\frac{\partial g\left(\vec x-\vec R(t)\right)}{\partial x_{i}}$ and the operator $L_{+}$ is defined by \eqref{Lplusedef}. For convenience, we denote $x_{i}=x^{i}$ and $R_{i}=R^{i}$.

The form of equations \eqref{eq1motion}, \eqref{eq2motion} suggests that in the leading order a solution for the fields $u$, $\rho$ and $v$ contains the terms $\sim\ddot R_{i}$, $\sim\dot R_{i}\dot R_{j}$. Thus, the time derivatives of these fields contain the terms with more than two derivatives in time $t$. Such terms were omitted in the derivation of equations \eqref{eq1motion}, \eqref{eq2motion}, thus, in the leading order we can also omit the terms $\dot u$, $\dot v$, $\ddot u$, $\ddot v$, $\ddot\rho$. It is clear that the radiation which is produced by the accelerated motion of the Q-ball, turns out to be neglected too. The equation for the field $v$ decouples in this case and we can simply set $v\equiv 0$. Thus, we arrive at the system of equations
\begin{equation}\label{urhoeqreduced}
L_{+}\begin{pmatrix}
u \\ \rho
\end{pmatrix}=\begin{pmatrix}
\sum\limits_{i=1}^{3}\left(\ddot{R_{i}}\partial_{i}f-2\omega^{2}f\ddot{R_{i}}x_{i}-\sum\limits_{j=1}^{3}
\dot{R_{i}}\dot{R_{j}}\partial_{i}\partial_{j}f-\omega^{2}f\dot{R_{i}}^{2}\right)-hf U
 \\ \frac{1}{2}\sum\limits_{i=1}^{3}\left(\ddot{R_{i}}\partial_{i}g-\sum\limits_{j=1}^{3}
\dot{R_{i}}\dot{R_{j}}\partial_{i}\partial_{j}g\right)
\end{pmatrix}.
\end{equation}

At the moment, we do not have an analytic solution even to simplified system of equations \eqref{urhoeqreduced}. However, using these equations we can get a very important consequence. Recall that \cite{Panin:2016ooo}
\begin{equation}\label{transmodes}
L_{+}\begin{pmatrix}
\partial_{i}f \\ \partial_{i}g
\end{pmatrix}=0,
\end{equation}
see also \eqref{transmode}. Multiplying \eqref{urhoeqreduced} by $\begin{pmatrix}
\partial_{k}f & \partial_{k}g \end{pmatrix}$, integrating the result over the space and using \eqref{transmodes}, we get
\begin{equation}\label{Newtlawprelim}
\ddot R_{k}\int\left((\partial_{k}f)^{2}+\frac{1}{2}(\partial_{k}g)^{2}+\omega^{2}f^{2}\right)d^{3}x+\frac{h}{2}\int f^{2}\partial_{k} U d^{3}x=0.
\end{equation}
Now let us make the following two steps. First, since the function $f(\vec x-\vec R(t))$ is localized in the vicinity of $\vec x=\vec R(t)$, for slowly varying $U(\vec x)$ we can write
\begin{equation}\label{intprelim}
\int f^{2}(\vec x-\vec R(t))\frac{\partial U(\vec x)}{\partial x_{k}} d^{3}x\approx \frac{\partial U(\vec R)}{\partial R_{k}}\int f^{2}(\vec x-\vec R(t))d^{3}x=\frac{I\sqrt{m^2-\omega^{2}}}{h^2}\,\frac{\partial U(\vec R)}{\partial R_{k}}.
\end{equation}
Second, using the spherical symmetry of the Q-ball solution with respect to the point $\vec x=\vec R(t)$, we can write
\begin{equation}\label{energyprelim}
\int\left((\partial_{k}f)^{2}+\frac{1}{2}(\partial_{k}g)^{2}+\omega^{2}f^{2}\right)d^{3}x=
\int\left(\frac{1}{3}\sum\limits_{i=1}^{3}(\partial_{i}f)^{2}+\frac{1}{6}\sum\limits_{i=1}^{3}(\partial_{i}g)^{2}+\omega^{2}f^{2}\right)d^{3}x.
\end{equation}
In paper \cite{Nugaev:2016uqd} it was shown that the relation
\begin{equation}\label{scaletr}
(\omega^{2}-m^{2})\int f^{2}\,d^3x+\int\sum\limits_{i=1}^{3}(\partial_{i}f)^{2}d^3x+\frac{1}{2}\int\sum\limits_{i=1}^{3}(\partial_{i}g)^{2}d^3x=0
\end{equation}
fulfills for Q-balls in the Wick--Cutkosky model. Substituting \eqref{scaletr} into \eqref{energyprelim} and using definition of the Q-ball rest energy \eqref{Q}, we arrive at
\begin{equation}\label{energyNlaw}
\int\left((\partial_{k}f)^{2}+\frac{1}{2}(\partial_{k}g)^{2}+\omega^{2}f^{2}\right)d^{3}x=
\frac{2\omega^{2}+m^{2}}{3}\int f^{2}\,d^{3}x=\frac{1}{2}E.
\end{equation}
Finally, using \eqref{intprelim} and \eqref{energyNlaw}, we obtain for \eqref{Newtlawprelim}
\begin{equation}\label{Nlaw}
E\ddot R_{k}=-\frac{I\sqrt{m^2-\omega^{2}}}{h}\,\frac{\partial U(\vec R)}{\partial R_{k}}.
\end{equation}
Using the definition of the Q-ball scalar charge (note that this charge is non-conserved in general)
\begin{equation}
Q_{SC}(\omega)=2m\int f^{2}\,d^{3}x,
\end{equation}
which was proposed in \cite{Nugaev:2016uqd} (for earlier discussion of the Q-ball scalar charge, see \cite{Levin:2010gp}), equation \eqref{Nlaw} can be rewritten as
\begin{equation}
E\ddot R_{k}=-\frac{h}{2m}Q_{SC}\frac{\partial U(\vec R)}{\partial R_{k}}.
\end{equation}
We see that, as expected, the Q-ball as a whole obeys the standard Newton law. However, the form of the Q-ball turns out to be modified during the accelerated motion, in the leading approximation this modification is determined, according to \eqref{urhoeqreduced}, by the components of the Q-ball acceleration $\ddot R_{i}$ and the products of the components of the Q-ball speed $\dot R_{i}\dot R_{j}$.

\section{Conclusion}
In the present paper, we discussed Q-balls in the Wick--Cutkosky model. Because of its simplicity, the model turns out to be a useful toy model for examining various properties of Q-balls, which can be inherent to Q-balls in other theories. In particular, we performed a detailed study of the Q-ball stability, including the stability with respect to small perturbations and spherically symmetric nonlinear evolution of the classically unstable Q-balls. The results of the analysis demonstrate that, depending on the characteristics of the Q-ball and on the form of the initial perturbation, the nonlinear evolution of the classically unstable Q-balls can lead to completely different outcomes. Namely, such Q-balls can spread into spherical waves, collapse or evolve into a classically stable Q-ball. Analogous evolution of classically unstable Q-balls may occur in other models admitting classically unstable Q-ball solutions.

In addition, we examined the behaviour of classically stable Q-balls in external fields. The key feature of the analysis is that this long-range external field is of the same nature as the one that forms the Q-ball itself. It is shown that, at least in the non-relativistic limit, the Q-ball as a whole obeys the Newton law. Meanwhile, the accelerated motion of the Q-ball modifies its form in accordance with its speed and acceleration.

We hope that the results presented in this paper can be useful for the further study of the Q-ball behaviour in different models.

\section*{Acknowledgements}
The authors are grateful to E.~Nugaev for valuable discussions. The work was supported by Grant No. 16-12-10494 of the Russian Science Foundation.

\section*{Appendix: energy density of one-field Q-balls with $\omega=0$ at $r=0$}
Let us take the standard action
\begin{equation}
S=\int\Bigl(\partial_\mu\phi^*\partial^\mu\phi-V(\phi^{*}\phi)\Bigr)d^4x
\end{equation}
with $V(\phi^{*}\phi)|_{\phi^{*}\phi=0}=0$. The equation of motion
\begin{equation}\label{eqmotapp}
\frac{\partial^{2}\phi}{\partial t^{2}}-\frac{\partial^{2}\phi}{\partial r^{2}}-\frac{2}{r}\frac{\partial\phi}{\partial r}+\frac{\partial V}{\partial(\phi^{*}\phi)}\phi=0
\end{equation}
is supposed to provide a Q-ball solution of the form $\phi(t,r)=e^{i\omega t}f(r)$. Suppose that there exists a Q-ball solution with $\omega=0$. Then, for $\omega=0$ the energy density at $r=0$ takes the form
\begin{equation}\label{ed0app}
\rho(0)=\left(\omega^{2}f^{2}+\left(\frac{df}{dr}\right)^{2}+V\right)\Biggl|_{\substack{r=0\\ \omega=0}}=V|_{r=0},
\end{equation}
because $\frac{df}{dr}\bigl|_{r=0}=0$. From equation of motion \eqref{eqmotapp} with $\omega=0$ it follows that
\begin{equation}\label{eqmotapp2}
-\frac{d^{2}f}{dr^{2}}-\frac{2}{r}\frac{df}{dr}+\frac{\partial V}{\partial(\phi^{*}\phi)}f=0.
\end{equation}
Let us multiply the latter equation by $\frac{df}{dr}$ and integrate the result over $r$ from $0$ to $\infty$. Using the facts that
\begin{equation}
\frac{\partial V}{\partial(\phi^{*}\phi)}f\frac{df}{dr}=\frac{1}{2}\frac{dV}{dr},
\end{equation}
$\frac{df}{dr}\bigl|_{r=0}=0$ and $\lim\limits_{r\to\infty}f(r)=0$, we get
\begin{equation}
V|_{r=0}=-\int\limits_{0}^{\infty}\frac{4}{r}\left(\frac{df}{dr}\right)^{2}dr<0.
\end{equation}
The latter means that, according to \eqref{ed0app}, $\rho(0)<0$ for $\omega=0$.

Now let us turn to $\frac{d^{2}\rho(r)}{dr^{2}}\bigl|_{r=0}$ for $\omega=0$. It is clear that for small $r$ the Q-ball profile $f(r)$ can be approximated as
\begin{equation}\label{fapproxApp}
f(r)\approx a-br^{2},
\end{equation}
where, without loss of generality, we can take $a>0$, $b>0$. With \eqref{fapproxApp}, for the energy density at $\omega=0$ we get
\begin{equation}\label{ED2derivApp}
\frac{d^{2}\rho(r)}{dr^{2}}\biggl|_{r=0}=\frac{d^{2}}{dr^{2}}\left(\left(\frac{df}{dr}\right)^{2}+V\right)\Biggl|_{r=0}=8b^{2}
-4ba\frac{\partial V}{\partial(\phi^{*}\phi)}\biggl|_{r=0}.
\end{equation}
Substituting \eqref{fapproxApp} into equation of motion \eqref{eqmotapp2}, for $r=0$ we obtain
\begin{equation}
a\frac{\partial V}{\partial(\phi^{*}\phi)}\biggl|_{r=0}=-6b.
\end{equation}
Substituting the latter result into \eqref{ED2derivApp}, finally we get
\begin{equation}
\frac{d^{2}\rho(r)}{dr^{2}}\biggl|_{r=0}=32b^{2}>0.
\end{equation}

\end{document}